\documentclass[aps,prd,twocolumn,superscriptaddress,preprintnumbers,floatfix,nofootinbib,notitlepage,showkeys,showpacs]{revtex4-1}

\usepackage[utf8]{inputenc}

\usepackage{graphicx}
\usepackage{hyperref}
\usepackage{latexsym}
\usepackage{amsmath}
\usepackage{amssymb}
\usepackage{bbm}
\usepackage{ulem}
\usepackage{pdfsync}
\usepackage{epsfig}
\usepackage{epstopdf}
\usepackage{subfigure}
\usepackage{color}
\usepackage{comment}
\usepackage{slashed}

\newcommand{\be}{\begin{equation}}
\newcommand{\ee}{\end{equation}}
\newcommand{\bea}{\begin{equation}\begin{aligned}}
\newcommand{\eea}{\end{aligned}\end{equation}}

\def\lsim{\mathrel{\raise.3ex\hbox{$<$\kern-.75em\lower1ex\hbox{$\sim$}}}}
\def\gsim{\mathrel{\raise.3ex\hbox{$>$\kern-.75em\lower1ex\hbox{$\sim$}}}}

\begin{document}

\title{Primordial black holes from inflaton and spectator field perturbations in a matter-dominated era}

\author{Bernard Carr}
\email{b.j.carr@qmul.ac.uk}
\affiliation{Astronomy Unit, Queen Mary University of London, \\ Mile End Road, London, E1 4NS, U.K.}
\author{Tommi Tenkanen}
\email{t.tenkanen@qmul.ac.uk}
\affiliation{Astronomy Unit, Queen Mary University of London, \\ Mile End Road, London, E1 4NS, U.K.}
\author{Ville Vaskonen}
\email{ville.vaskonen@kbfi.ee}
\affiliation{National Institute of Chemical Physics and Biophysics, \\ R\"avala 10, 10143 Tallinn, Estonia}

\begin{abstract}
We study production of primordial black holes (PBHs) during an early matter-dominated phase. As a source of perturbations, we consider either an inflaton field with a running spectral index or a spectator field that has a blue spectrum and thus provides a significant contribution to PBH production at small scales. First, we identify the region of the parameter space where a significant fraction of the observed dark matter can be produced, taking into account all current PBH constraints. Then, we present constraints on the amplitude and spectral index of the spectator field as a function of the reheating temperature. We also derive constraints on the running of the inflaton spectral index, ${\rm d}n/{\rm d}{\rm ln}k \lesssim 0.001$, which are comparable to those from the Planck satellite for a scenario where the spectator field is absent.
\end{abstract}

\maketitle


\section{Introduction}

Recently, primordial black holes (PBHs) have received much attention~\cite{Carr:2016drx,Green:2016xgy,Harada:2016mhb,Kuhnel:2017pwq,Cotner:2016cvr,Georg:2017mqk,Carr:2017jsz,Emami:2017fiy,Kuhnel:2017bvu,Green:2017qoa}. In particular, there have been many studies of PBH constraints on the primordial power spectrum associated with the inflationary scenario since this provides an efficient way to probe different models of inflation and reheating~\cite{Carr:1994ar,Green:1997sz,Bringmann:2001yp,Josan:2009qn,Khlopov:2004tn,Bringmann:2011ut,Garcia-Bellido:2017mdw,Kannike:2017bxn}. There has also been interest in constraints on PBHs formed during an early matter-dominated era, when PBHs form more easily. This was originally considered in Refs.~\cite{Khlopov:1980mg,Polnarev:1986bi} and more recently in Refs.~\cite{Harada:2016mhb,Georg:2016yxa,Cotner:2016cvr,Georg:2017mqk,Carr:2017jsz}.

In this work, we use the most up-to-date astrophysical and cosmological constraints to consider whether PBHs produced during an early matter-dominated phase can constitute all the dark matter (DM) and what PBHs can tell us about the curvature perturbation spectrum on small scales. We will first assume that only one component, the inflaton field, determines the perturbation spectrum on all scales and take the amplitude and spectral index of its power spectrum to be given by the best fit to the Planck data. Studying PBH production provides important constraints on different models of inflation and reheating. To demonstrate this, we will allow a positive value for the running of the inflaton spectral index, corresponding to a `blue' spectrum, and show that the constraints on this can be comparable to those from the Planck data. For earlier studies of PBH formation in scenarios with a running spectral index, see Refs.~\cite{Alabidi:2012ex,Alabidi:2013wtp}.

Next we will assume that there are two scalar fields, which together determine the perturbation spectrum: the inflaton field, which gives the dominant component to the curvature spectrum at {\it large} scales, and a spectator field which gives the dominant component  at {\it small} scales. By the term `spectator field' we mean a scalar field which was energetically subdominant during cosmic inflation and played no role in driving or ending inflation. If such a scalar field were sufficiently light, it would have acquired a spectrum of perturbations uncorrelated with perturbations in the inflaton sector and with a potentially large amplitude.

The scenario is well motivated, as spectator fields are a generic ingredient if one goes beyond the Standard Model (SM) of particle physics. They can affect the physics of the early Universe in a number of ways, including the generation of the curvature power spectrum~\cite{Lyth:2001nq,Enqvist:2001zp,Moroi:2001ct,Dvali:2003em}, matter-antimatter asymmetry~\cite{Kusenko:2014lra} and dark matter~\cite{Nurmi:2015ema,Markkanen:2015xuw}. A well-known example is the SM Higgs field, whose cosmological implications have been studied in detail in 
a number of works~\cite{DeSimone:2012qr,Enqvist:2013kaa,Herranen:2014cua,Herranen:2015ima,Espinosa:2015qea}. As another example, the generation of PBHs for a specific type of spectator, a curvaton field~\cite{Lyth:2001nq,Enqvist:2001zp,Moroi:2001ct}, has been considered in Refs.~\cite{Kawasaki:2012wr,Firouzjahi:2012iz,Kohri:2012yw,Young:2013oia,Carr:2016drx}.

Spectator fields can also produce an early matter-dominated period in the early Universe. For example, if they are very massive and sufficiently long-lived, they may come to dominate the energy density before decaying into SM radiation. For two recent examples of such a scenario, see Refs.~\cite{Berlin:2016vnh,Tenkanen:2016jic}. An early matter-dominated period can also arise in scenarios where there are no additional fields which dominate the energy density, as in the case of slow reheating \cite{Carr:1994ar,Allahverdi:2010xz}. For a recent study of PBH formation in such a scenario, see Ref.~\cite{Hidalgo:2017dfp}.  In this paper our treatment is very general in that we do not specify the cause of the matter-dominated phase.

Understanding the properties of spectator fields can have far-reaching implications for different phenomena, including those unrelated to PBH formation. Placing constraints on the DM abundance, the duration of the early matter-dominated phase and the reheating temperature is important in addressing long-standing questions in physics, such as the origin of the DM and matter-antimatter asymmetry. In this paper, we use the most recent constraints on PBHs to present new limits on PBH DM, running of the inflaton spectral index, and spectral features of generic spectator fields.

The paper is organised as follows. In Section~\ref{spectrum} we set up the scenario and discuss how the the curvature power spectrum arises at different scales and relates to the free parameters of the model. In Section~\ref{formation} we discuss the formation of PBHs during an early matter-dominated era. In Section~\ref{constraints} we consider their role as dark matter and in Section~\ref{results} we place associated constraints on the parameters of the model. We summarise our key conclusions in Section~\ref{conclusions}.

\section{The curvature power spectrum}
\label{spectrum}

As is well known, perturbations crossing the horizon during a matter-dominated phase grow linearly with the scale factor $a$ until the end of matter domination. This opens up new possibilities for the formation of PBHs, as their production rate can be vastly enhanced compared to the usual radiation-dominated scenario~\cite{Carr:1974nx, Khlopov:1980mg,Polnarev:1986bi}. If this happened prior to big bang nucleosynthesis (BBN), one can ask whether these PBHs could constitute a significant fraction of the observed DM abundance or whether they could probe the curvature perturbations at scales which are otherwise unobservable{\footnote{The PBHs could themselves be a source of large-scale fluctuations as a result of the Poisson fluctuations in their number density but these are distinct from the primordial fluctuations.}.  

As an example of such a scenario, one can consider a scalar field which dominated the energy density of the very early Universe and subsequently decayed into SM radiation~\cite{Berlin:2016vnh,Tenkanen:2016jic}. We will also consider a scenario where a prolonged period of reheating gives an effectively matter-dominated period even if there were no additional fields which dominated the energy density of the Universe. However, our conclusions do not depend on what causes the matter domination. 

Let us assume that the total curvature power spectrum is given by the sum of the perturbations produced from the inflaton $\varphi$ and the spectator $s$: 
\be
\label{PRs}
\mathcal{P}_\mathcal{R}(k) = \langle |\mathcal{R}_k|^2\rangle = \mathcal{P}_{\mathcal{R},\varphi}(k) + \mathcal{P}_{\mathcal{R},s}(k).
\ee 
The first term is assumed to dominate on large scales (small Fourier mode $k$) and the second on small scales (large $k$). Here $\mathcal{R}_k$ is the comoving curvature perturbation, which for modes $k$ outside the horizon is related to the metric perturbation $\Phi(k)$ via
\be
\label{Req}
\mathcal{R}_k = - \frac{(5+3w) \Phi(k)}{3+3w} \, ,
\ee
where $w$ is the equation of state parameter, $p = w \rho$. Throughout this paper we use the longitudinal gauge,
\be
{\rm d}s^2 = -(1+2\Phi){\rm d}t^2 + a(t)^2(1-2\Phi){\rm d}x^2 \, ,
\ee
where we assume vanishing anisotropic stress and use natural units with $\hbar = c =1$.

The inflaton perturbations are assumed to produce a nearly flat spectrum at small $k$ in accordance with the observations of the Cosmic Microwave Background (CMB):
\be
\mathcal{P}_{\mathcal{R},\varphi}(k) = A \left(\frac{k}{k_*}\right)^{n-1 + \frac12\alpha\ln\left(\frac{k}{k_*}\right)},
\ee 
where $k_* = 0.05$ Mpc$\mbox{}^{-1}$ is the Planck pivot scale, $\log(10^{10} A) = 3.062\pm0.029$ is the power spectrum amplitude, and $n = 0.968\pm0.006$ is the scalar spectral index~\cite{Ade:2015lrj}. We also allow for the running of the spectral index, the Planck measurements indicating~\cite{Ade:2015lrj} 
\be
\label{alpha}
 \alpha \equiv \frac{{\rm d}n}{{\rm d}{\rm ln}k} = -0.0033\pm0.0074 \,.
\ee
We will show below that the current constraints on running from PBH formation can be comparable to this in certain circumstances. For simplicity, we assume the running of the running is negligible.

At large $k$, perturbations in the $s$ field can dominate the power spectrum and generate sufficiently large metric perturbations to produce a significant PBH abundance during an early matter-dominated phase. Let us assume that the $s$ perturbation spectrum is given by
\be \label{spsp}
\mathcal{P}_{\mathcal{R},s}(k) = A_s \left(\frac{k}{k_*}\right)^{n_s - 1 + \frac12\alpha_s\ln\left(\frac{k}{k_*}\right)},
\ee
where $A_s$ is the amplitude of the $s$ power spectrum, $n_s$ is the $s$ spectral index, and we again allow for a non-zero running, $\alpha_s\equiv {\rm d}n_s/{\rm d}{\rm ln}k$. Further, we assume that $A_s\ll A$, $n_s>1$ and $\alpha_s\leq 0$. The total power spectrum is illustrated in Fig.~\ref{pstot}.

\begin{figure}
\includegraphics[width=.45\textwidth]{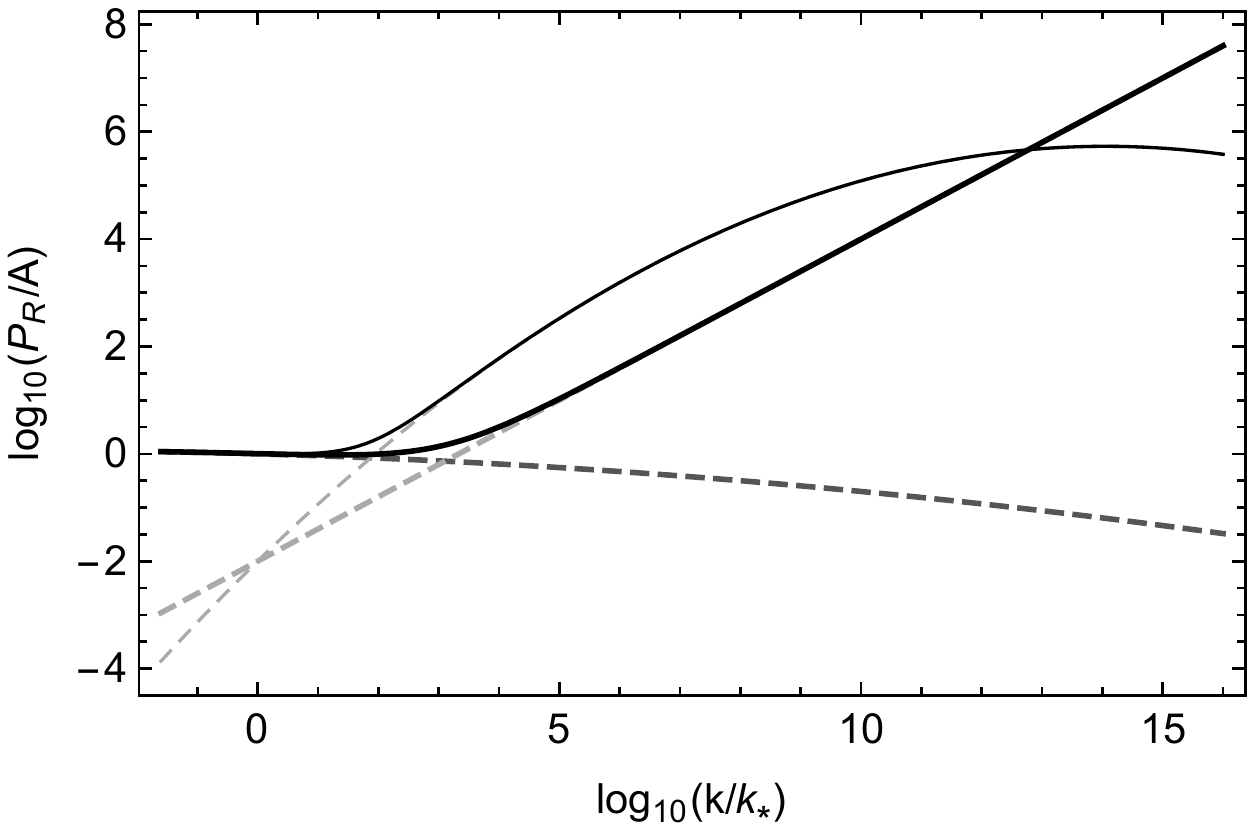}
\caption{Contributions of the inflaton (black dashed line) and the spectator field (gray dashed line) to the total curvature power spectrum (black solid line). The thick and thin lines correspond to $(n_s,\alpha_s,A_s/A)=(1.6,0,0.01)$ and $(n_s,\alpha_s,A_s/A)=(2.1,-0.034,0.01)$, respectively.}
\label{pstot}
\end{figure}

The scenario involves five extra free parameters: the reheating temperature\footnote{By reheating temperature we mean the temperature of the baryon--photon fluid at the time it becomes the dominant energy density component. This temperature may be unrelated to the energy scale of cosmic inflation. } $T_{\rm reh}$, which relates to the $s$ decay width via
\be
\Gamma = \sqrt{\frac{4\pi^3}{45}g_*(T_{\rm reh})} \,  \frac{T_{\rm reh}^2}{M_{\rm P}} \, ,
\ee
where $M_{\rm P} = \sqrt{1/G}$ is the Planck mass; the duration of the early matter-dominated phase; the amplitude of the perturbation spectrum of $s$; the corresponding spectral index and its running. For successful BBN the reheating temperature has to satisfy $T_{\rm reh}\gsim 4$ MeV~\cite{Kawasaki:2000en,Hannestad:2004px,Ichikawa:2005vw,DeBernardis:2008zz}.

Before presenting the PBH yield in this scenario, let us briefly motivate these parameter choices and discuss some possible issues. First, the spectral index of $s$ is given by
\be \label{spectralindex}
n_s - 1 = 3-3\sqrt{1-\frac49\lambda} \, ,
\ee
where $\lambda$ is a parameter controlling the effective mass of the spectator field during inflation, 
\be
V_s = \frac{\lambda}{2} H^2s^2 \, ,
\ee
$H$ being the Hubble scale. In the case of a non-minimal coupling to the curvature scalar, $\xi s^2 R$ appearing in the total Lagrangian, one has $\lambda = 12\xi$ during a de Sitter inflationary phase, so $\lambda\sim 1$ is a natural choice if one takes $\xi\lesssim 1$. The spectral index can therefore be large, $4 \geq n_s \gtrsim 1$. This case is not only of interest for PBH formation because this type of non-minimal gravitational coupling has been shown to arise due to quantum corrections in a curved spacetime even if it is initially set to zero~\cite{Freedman:1974gs}. Indeed, for the SM Higgs field, a coupling in the range $\mathcal{O}(10^{-2})\lesssim \xi \lesssim \mathcal{O}(10)$ is necessary to avoid catastrophic collapse of the Universe due to the SM vacuum instability \cite{Herranen:2014cua, Herranen:2015ima,Espinosa:2015qea,Kohri:2016wof,Ema:2016kpf}. However, in this paper our approach is more general, in the sense that both during and after inflation we specify neither the potential governing the dynamics of the spectator field nor the inflaton field itself, but treat $n_s$ and $\alpha_s$ as free parameters.

Second, because the contribution of the spectator field to $\mathcal{P}_{\mathcal{R}}$ is assumed to be negligible at small $k$, the associated isocurvature mode is unobservable in the CMB or large-scale structure measurements. Also the possible non-Gaussianity arising in multifield models, which may affect constraints on PBH production in certain cases~\cite{Young:2013oia,Tada:2015noa,Young:2015kda}, is not an issue here. This is because PBH formation does not occur on the tail of the distribution of curvature fluctuations in the matter-dominated case, as we now discuss.

\section{PBH formation}
\label{formation}

An overdense region on scale $k$ in the early Universe is characterised by its density contrast $\delta = (\rho-\bar\rho)/\bar\rho$. In a radiation-dominated period an overdense region can collapse against the pressure to form a black hole if the density contrast at horizon crossing, $k=aH$, exceeds some critical value which was originally taken to be $\delta_c \approx 1/3$~\cite{Carr:1975qj}. Later numerical studies have refined this estimate~\cite{Niemeyer:1999ak,Musco:2004ak,Musco:2008hv,Musco:2012au,Harada:2013epa} and here we adopt the value $\delta_c=0.45$ found in Ref.~\cite{Musco:2004ak}. 

We first illustrate PBH formation in the usual radiation-dominated case, assuming that the fluctuations at horizon crossing are Gaussian with variance $\sigma^2(k)=4\langle |\Phi(k)|^2\rangle/9$. The fraction of the Universe collapsing into PBHs of mass $M$ is then \cite{Carr:1975qj}
\bea
\label{betarad}
\beta(M) &=  \frac{2}{\sqrt{2\pi}\sigma(M)}\int_{\delta_c}^\infty {\rm d}\delta\, \exp\left(-\frac{\delta^2}{\sigma(M)^2}\right) \\
&= {\rm Erfc}\left( \frac{\delta_c}{\sqrt{2}\sigma(M)} \right) \,,
\eea
where ${\rm Erfc}$ is the complementary error function. The horizon mass $M$ and the scale $k$ are related by $M\propto 1/k$. The production of PBHs is exponentially suppressed because they only form from the tail of the density fluctuation distribution. This also means that any region collapsing to a PBH is likely to be nearly spherically symmetric~\cite{1970Ap......6..320D,1986ApJ...304...15B}.

One would expect the width of the PBH mass distribution to be at least $\Delta M \sim M$ and in many scenarios it would be much more extended than this~\cite{Carr:2016drx}. Thus, following Ref.~\cite{Carr:2017jsz}, we define the PBH mass function as
\be
\psi(M)\equiv  \frac{1}{\rho_{\rm DM}} \frac{{\rm d}\rho_{\rm PBH}(M)}{{\rm d}M}\,, 
\ee
where $\rho_{\rm DM}$ is the observed DM abundance, so that the fraction of the DM density in PBHs in the mass interval $(M,M+{\rm d}M)$ is $\psi(M){\rm d}M$. 

Since the PBH and radiation densities scale as $a^{-3}$ and $a^{-4}$, respectively, there is a simple relationship between $\beta(M)$ and $\psi(M)$ at the present epoch:
\be
\psi(M){\rm d}M = \frac{a_{\rm eq}}{a(M)} \frac{\beta(M)}{M} {\rm d}M\,, 
\ee
where $a(M)$ corresponds to the time when PBHs of mass $M$ are formed and $a_{\rm eq} = a(z=3365)$ to the time of matter-radiation equality (i.e. we assume there is no matter-dominated period before then). Since PBHs are expected to have roughly the horizon mass at formation,
\be \label{psi}
M \approx 
\frac{4 \pi\rho_{\rm tot}}{3H^3} = \frac{M_{\rm P}^2}{2H}\,,
\ee 
this gives
\be \label{frelrad}
\psi(M){\rm d}M \approx 5\times10^{8}\left(\frac{M_\odot}{M}\right)^{1/2} \frac{\beta(M)}{M} \, {\rm d}M \,,
\ee
where $\rho_{\rm tot}=3H^2M_{\rm P}^2/(8\pi)$ is the total energy density.

In an early matter-dominated era, we have $p=0$ (or $w=0$), so the situation is different in several respects. First, the density contrast grows linearly, $\delta \propto a$, after horizon crossing, whereas in a radiation-dominated era it stops growing because of the effect of pressure below the Jeans length. This means that a PBH can be formed from a density contrast which is much smaller than $\delta_c$ at the horizon crossing. However, it also means that the region must collapse  a lot before forming a PBH, so any appreciable initial asphericity will be amplified and this could prevent black hole formation. The probability of PBH formation therefore depends upon the fraction of regions which are sufficiently spherically symmetric. According to the analysis of Ref.~\cite{Harada:2016mhb}, this is given by
\be
\label{betamd}
\beta(M) \simeq 0.056\, \sigma(M)^5 \,.
\ee
The factor $\sigma(M)^5$ was first derived
in Refs.~\cite{Khlopov:1980mg,Polnarev:1986bi} and these papers also included an extra suppression factor $\sigma(M)^{3/2}$ in order to take into account inhomogeneity effects, which might decrease the probability of PBH formation\footnote{Note that in Refs.~\cite{Khlopov:1980mg,Polnarev:1986bi} the variance at the beginning of the matter-dominated era is denoted by $\delta^2$, so an extra factor $(M/M_0)^{13/3}$ appears, where $M_0$ is the horizon mass at the beginning of matter-dominance, due to the growth of the density perturbation in the synchronous gauge. In our case, $\sigma(M)$ is the value at horizon-crossing, so this factor is absent.}. As discussed in Ref.~\cite{Harada:2016mhb}, there is some uncertainty about this factor, so in Sec.~\ref{results} we show the results with and without the factor of $\sigma(M)^{3/2}$.

Second, the relationships \eqref{psi} and \eqref{frelrad} no longer apply since the ratio $\rho_{\rm PBH}/\rho_{\rm tot}$ remains constant during the matter-dominated era.  For PBHs forming in this period, we can therefore identify the fraction of the Universe in PBHs at formation with the fraction at the reheating epoch, so the PBH DM fraction in the mass interval $(M,M+{\rm d}M)$ at the present epoch is
\bea \label{frelmd}
\psi(M){\rm d}M &= \frac{a_{\rm eq}}{a_{\rm reh}} \frac{\beta(M)}{M}\,  {\rm d}M \\
&\simeq 7\times10^{27}\left(\frac{\Gamma}{M_{\rm P}}\right)^{1/2} \frac{\beta(M)}{M} {\rm d}M \,,
\eea
and there is no $M^{1/2}$ factor in the relationship between $\beta(M)/M$ and $\psi(M)$.

The smallest and largest scales which become non-linear during the matter-dominated era (i.e. such that $\delta\simeq \sigma$ reaches $1$ when $a \leq a_{\rm reh}$) are given by
\be \label{masses}
M_{\rm min} = \frac{M_{\rm P}^2}{2\Gamma} \left(\frac{a_{\rm md}}{a_{\rm reh}}\right)^{3/2}\,, \quad M_{\rm max} = \frac{M_{\rm P}^2}{2\Gamma} \, \sigma_{\rm max}^{3/2} \,.
\ee
Here $\sigma_{\rm max}^2$ is the variance of the density contrast distribution at the time the scale related to $M_{\rm max}$ enters the horizon and $a_{\rm md}$ is the scale factor at the beginning of matter dominance. The first expression in Eq.~\eqref{masses}  is the horizon mass at this time and the second expression is the horizon mass at the epoch when the regions which bind at the end of the matter dominance, $a=a_{\rm reh}$, enter the horizon. 

For the case in which the spectator field perturbations dominate over the ones of the inflaton field, the variance of the $s$ density contrast corresponding to the mass $M_{\rm max}$ can be solved from the equation
\be
\sigma_{\rm max}^2 = \left(\frac25\right)^2 A_s \left(\frac{k_{\rm max}}{k_*}\right)^{n_s-1+\frac12\ln\left(\frac{k_{\rm max}}{k_*}\right)} \,,
\ee
which follows from Eqs \eqref{PRs}, \eqref{Req}, and \eqref{spsp}. Here $k_{\rm max}=k_{\rm reh}/\sqrt{\sigma_{\rm max}}$ is the mode corresponding to the mass $M_{\rm max}$, where we define $k_{\rm reh}$ as the mode which re-enters horizon at the reheating temperature, $k_{\rm reh} = a_{\rm reh} \Gamma$. For the power spectrum~\eqref{spsp} Eq.~\eqref{betamd} then gives
\bea \label{betamd2}
\beta(M) = &  5.7\times10^{-4} A_s^{5/2} \left(\frac{M}{M_c}\right)^{-5(n_s-1)/6}  \\ &\times\exp\left[ \frac{5}{36}\alpha_s\ln^2\left(\frac{M}{M_c}\right) \right] \,,
\eea
where
\be
M_c = \left(\frac{k_{\rm reh}}{k_*}\right)^3\frac{M_{\rm P}^2}{2\Gamma} \, .
\ee
A similar result can be derived for the case in which the spectator field is absent and the dominant contribution to the power spectrum is due to the inflaton field.

\section{PBHs as dark matter}
\label{constraints}

The PBH mass function produced during the early matter dominance is given by Eq.~\eqref{frelmd} with the the minimum and maximum masses given by Eq.~\eqref{masses}:
\be \label{mf}
\psi(M) = \begin{cases} 
\frac{a_{\rm eq}}{a_{\rm reh}} \frac{\beta(M)}{M}  & (M_{\rm min}<M<M_{\rm max}) \\
0   & ({\rm otherwise}) \, .
\end{cases}
\ee
The function $\beta(M)$ is given by Eq.~\eqref{betamd2}. The fraction of DM consisting of PBHs, $f=\rho_{\rm PBH}/\rho_{\rm DM}$, is then obtained by integrating the PBH mass function:
\be
\label{totalenergy}
f = \int_{M_{\rm min}}^{M_{\rm max}} \psi(M) {\rm d}M \,.
\ee

\begin{figure}
\begin{center}
\includegraphics[width=.45\textwidth]{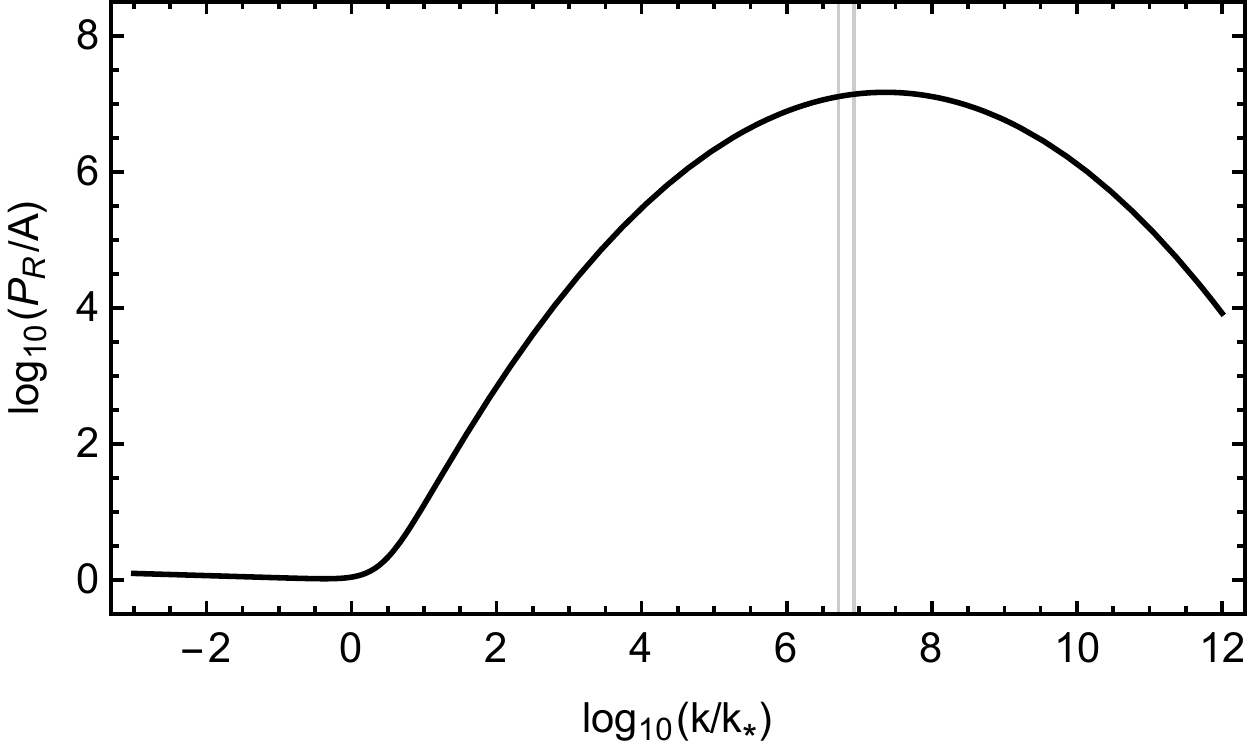} \\ \vspace{3mm}
\includegraphics[width=.45\textwidth]{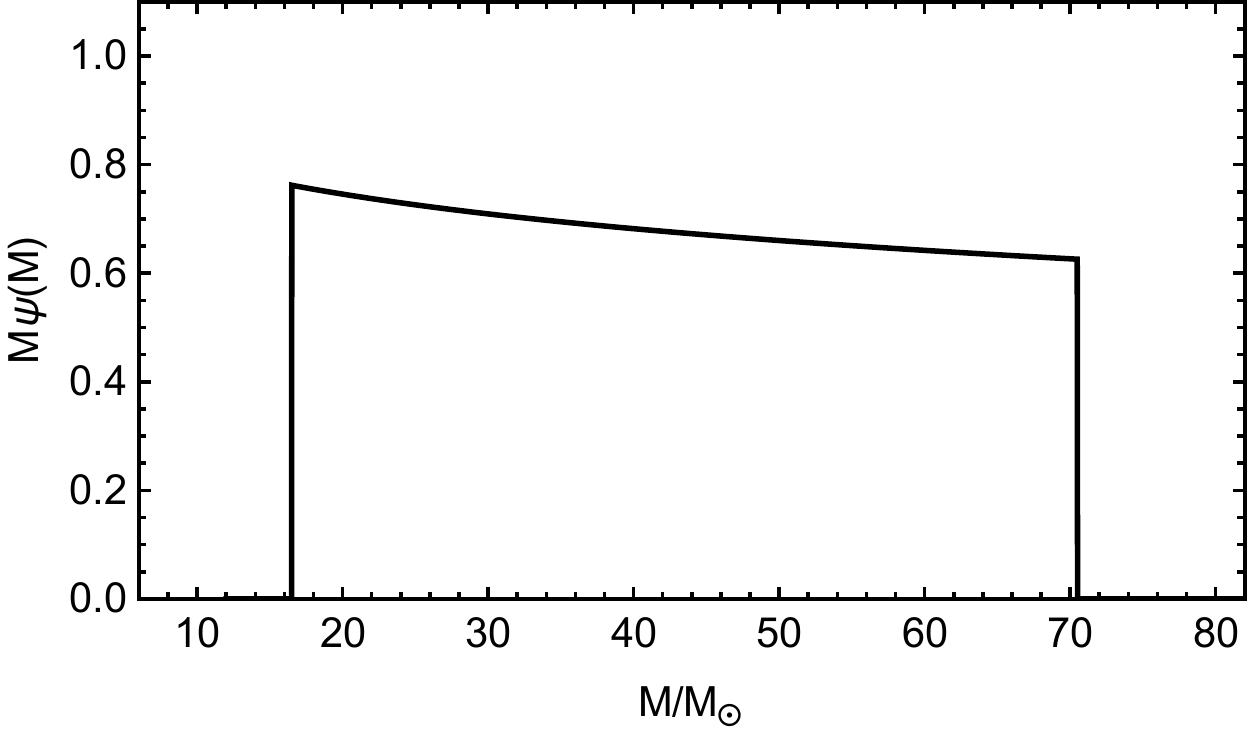}
\caption{The upper panel shows the power spectrum for $n_s=3.22$, $\alpha_s=-0.131$, $A/A_s=0.1$, $T_{\rm reh}=6$\,MeV and $a_{\rm reh}/a_{\rm md}=40$. The lower panel shows the corresponding PBH mass function. The smallest and largest modes which produce PBHs during the matter-dominated era are indicated in the upper 
panel by the vertical lines.}
\label{ex}
\end{center}
\end{figure}

If $\alpha_s$ is small, then the mass function has a power-law form with cut-offs at $M_{\rm min}$ and $M_{\rm max}$. In Fig.~\ref{ex} the mass function~\eqref{mf} is shown for the parameters given in the caption. The most important difference in the mass function produced during matter-dominance compared to the one produced during radiation-dominance with a similar power spectrum is the cut-offs arising from the finite duration of the matter dominance. This allows for larger PBH abundance, in better agreement with the constraints.

The PBH DM scenario is very strongly constrained. The main constraints arise from evaporation, lensing and accretion effects; see Refs~\cite{Carr:2016drx, Carr:2017jsz} for a recent summary. The constraints for a monochromatic PBH mass function are shown in Fig.~\ref{constr}. For the evaporation constraint we take $\epsilon=0.2$~\cite{Carr:2009jm} where $\epsilon$ gives the slope of the extragalactic $\gamma$-ray background; for the neutron-star capture constraint, we assume a DM density $\rho_{\rm DM} = 2\times10^3$\,GeV${\rm cm}^{-3}$ in the core of globular clusters~\cite{Capela:2013yf}; for the Planck accretion constraint we show the most conservative bound~\cite{Ali-Haimoud:2016mbv}. PBHs with $M< M_* \equiv 4 \times 10^{14}{\rm g} = 10^{-18.6}M_{\odot}$ (i.e. left of the vertical dashed line in Fig.~\ref{constr}) have evaporated by now~\cite{Hawking:1974rv}. The strongest constraints on their abundance come from BBN~\cite{Carr:2009jm}  and anisotropies in the Cosmic Microwave Background (CMB);  the latter arise because heating associated with the evaporation of PBHs between recombination and reionization would dampen small-scale anisotropies, contrary to observation \cite{Carr:2009jm}. 

Also discussed in Ref.~\cite{Carr:2016drx} are various dynamical constraints, including those associated with the Poisson fluctuations in the PBH number density \cite{Afshordi:2003zb}. These limits are important for large  $M$ but usually depend upon additional and possibly contentious astrophysical assumptions. They are therefore shown by dashed lines in Fig.~\ref{constr}, the Poisson limit being omitted because it is weaker than the others. On large mass scales there are also constraints from limits on the $\mu$-distortions in the CMB generated by the damping of fluctuations in the period before decoupling. However, these only correspond to PBH limits if one assumes some relationship between the fluctuations and PBH formation \cite{Nakama:2016kfq}. 

\begin{figure}
\begin{center}
\includegraphics[width=.45\textwidth]{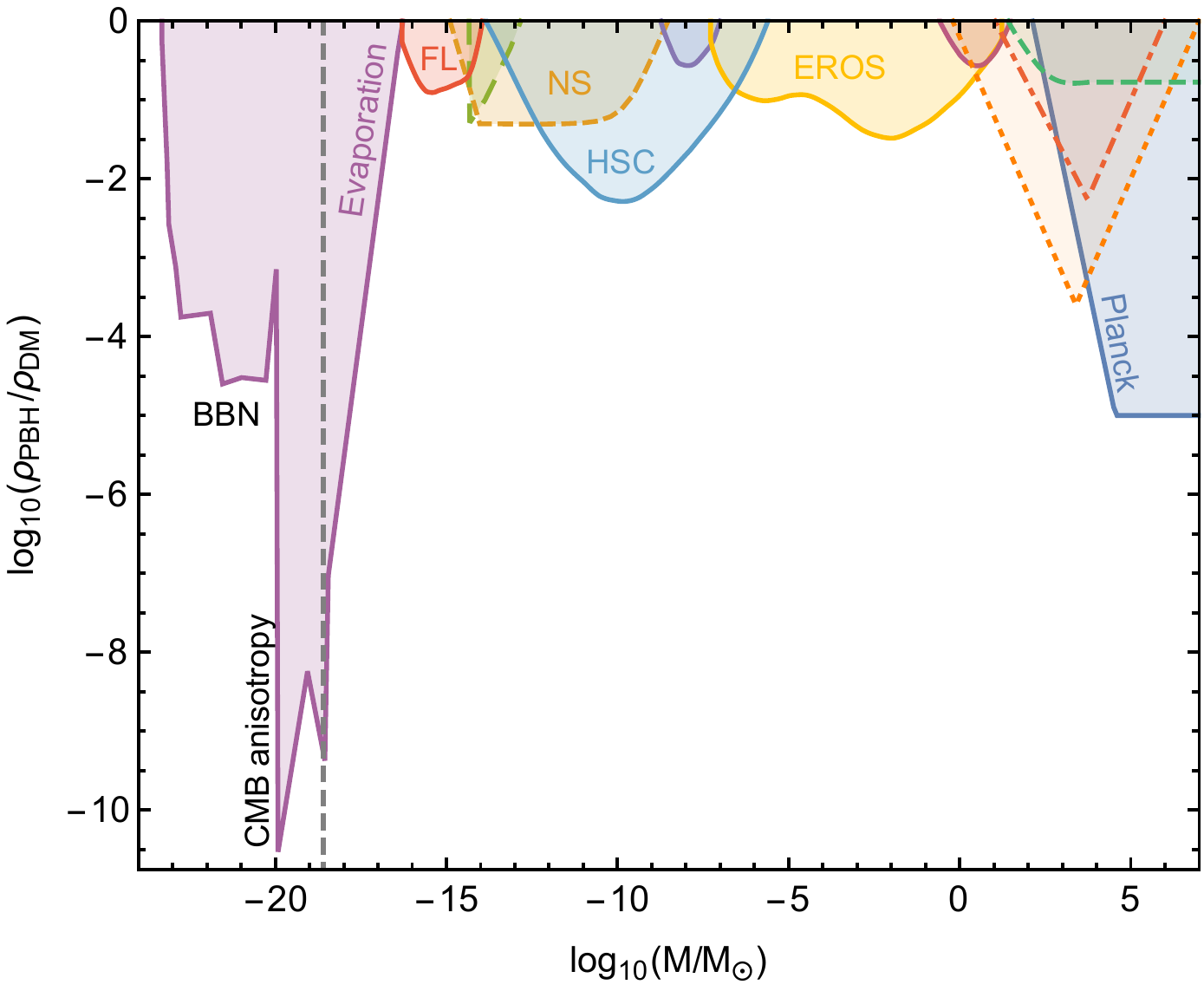}
\caption{The lines show the different constraint for a monochromatic PBH mass function. The purple region on the left is excluded by evaporation~\cite{Carr:2009jm}, the red region by femto-lensing of gamma-ray bursts~\cite{Barnacka:2012bm}, the brown region by neutron-star capture~\cite{Capela:2013yf}, the green region by white dwarf explosions~\cite{Graham:2015apa}, the blue, yellow and purple regions by microlensing results from Subaru~\cite{Niikura:2017zjd}, EROS~\cite{Tisserand:2006zx}, and MACHO~\cite{Allsman:2000kg} respectively, and the dark blue region by Planck~\cite{Ali-Haimoud:2016mbv}. The regions to the right of the dashed lines are excluded by survival of a stars in Segue I~\cite{Koushiappas:2017chw} and Eridanus II~\cite{Brandt:2016aco}, and distribution of wide binaries~\cite{Monroy-Rodriguez:2014ula}. PBHs to the left of the gray vertical line have evaporated before today.}
\label{constr}
\end{center}
\end{figure}

Following the method introduced in Ref.~\cite{Carr:2017jsz}, the constraints for a monochromatic mass function can straightforwardly be adapted for extended mass functions. Each constraint can be expressed in the form
\be \label{ineq}
\int {\rm d}M \frac{\psi(M)}{f_{\rm max}(M)} \leq 1\, ,
\ee
where $f_{\rm max}(M)$ is the constraint for monochromatic mass function, shown in Fig.~\ref{constr}. As shown in Ref.~\cite{Carr:2017jsz}, the wider the PBH mass function, the smaller the allowed PBH abundance. If the dynamical constraints are neglected}, the observed DM abundance can be obtained in the region $M\sim 10M_{\odot}$ with a sufficiently narrow PBH mass function. 

In our scenario the width of the PBH mass function depends not only on $n_s$ but also $\alpha_s$. More importantly, even if the top of the power spectrum is very wide, the  mass function can be narrow if the duration of matter dominance is short and close to the time when the top of the power spectrum enters the horizon, so that PBH production during any radiation-dominated era before or after the matter-dominated period is negligible. For example, choosing the parameters used in Fig.~\ref{ex} gives the observed DM abundance and the resulting mass function is nearly flat from $M_{\rm min}=16.5M_{\odot}$ to $M_{\rm max}=72.2M_{\odot}$. This mass function is in agreement with the lensing and accretion constraints but not with the dynamical constraints from Eridanus II, Segue I and wide binaries.\footnote{The accretion constraint is subject to large uncertainties, including its effect on the thermal history of the Universe. The most recent analysis~\cite{Poulin:2017bwe}, which appeared after the first version of this paper, claims much stronger constraints than the ones from Ref.~\cite{Ali-Haimoud:2016mbv} and would close the PBH DM window around $M \sim 10M_\odot$ even if the dynamical constraints were neglected.}

\begin{figure}
\begin{center}
\includegraphics[width=.45\textwidth]{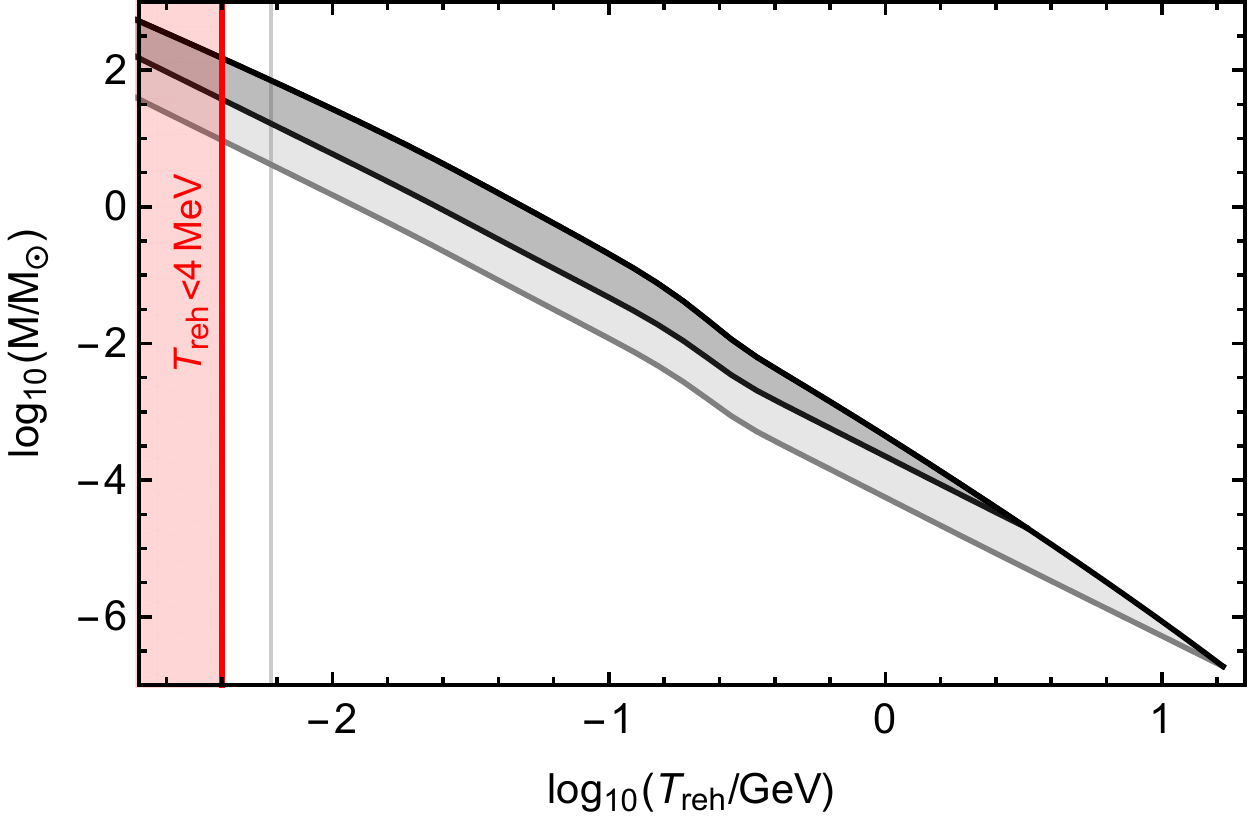} \\ \vspace{3mm}
\includegraphics[width=.45\textwidth]{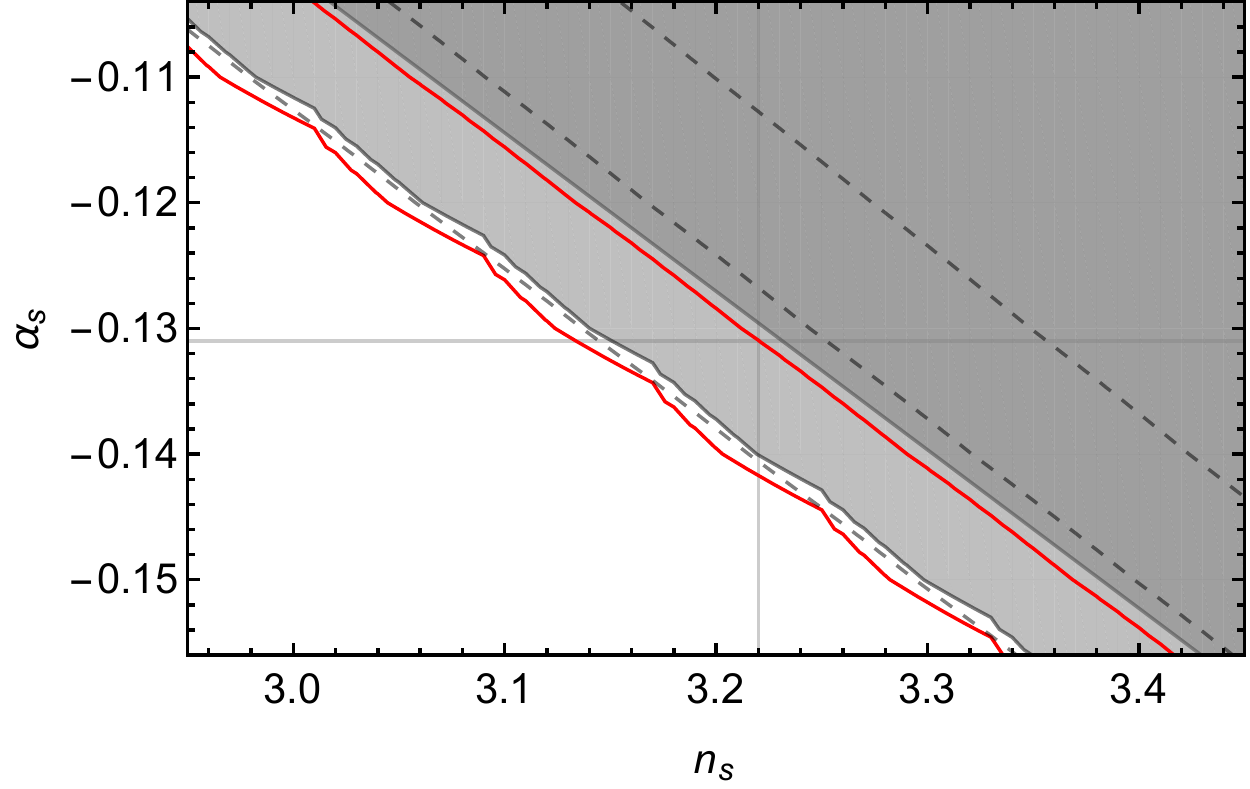}
\caption{Top panel: Minimum and maximum PBH masses. Black and gray lines correspond to $a_{\rm reh}/a_{\rm md}=40$ and $a_{\rm reh}/a_{\rm md}=100$, respectively. The red region is excluded by the BBN constraint on the reheating temperature. Bottom panel: The red contours correspond to 100\% (upper) and 1\% (lower) of PBH DM. The dark and light gray regions are excluded without and with dynamical constraints, respectively. The dashed contours from left to right correspond to $M_{\rm max}=30M_{\odot}$, $100M_{\odot}$ and $300M_{\odot}$. In both plots the values used in Fig.~\ref{ex} are depicted by the solid gray lines and other parameters are fixed to the same values.}
\label{ex2}
\end{center}
\end{figure}

In Fig.~\ref{ex2} the effects of changing parameters around the ones used in the previous example are shown. First, in the top plot, the gray regions show the mass range in which PBHs are formed during matter dominance. Obtaining PBH mass $M\sim 10M_{\odot}$ requires relatively low reheating temperature, $T_{\rm reh}\sim 10$ MeV. The bottom panel shows how the PBH abundance, the constraints and the maximum PBH mass change as a function of  $n_s$ and $\alpha_s$.

\section{Constraints on scalar field spectra}
\label{results}

For a monochromatic PBH mass function, the PBH constraints can be converted into an upper limit on the amplitude of the power spectrum at a given scale $k$. If the scale $k$ re-enters the horizon during a matter-dominated era, the upper bound on $\mathcal{P}_{\mathcal R}(k)$ is obtained from
\be
a_{\rm reh}^3 \rho_{\rm tot}(a_{\rm reh}) \beta\left(\sigma =  \frac{2}{5}\sqrt{\mathcal{P}_{\mathcal R}(k)}\right) < \rho_{\rm PBH}^{\rm max}\,,
\ee 
where $\beta$ is given by Eq.~\eqref{betamd} and the maximum PBH energy density today, $\rho_{\rm PBH}^{\rm max}$, can be read off from Fig.~\ref{constr}. If the mode $k$ re-enters the horizon during a radiation-dominated era, the constraint on $\mathcal{P}_{\mathcal R}(k)$ is obtained from
\be
a(M)^3 \rho_{\rm tot}(a(M)) \beta\left(\sigma = \frac{4}{9}\sqrt{\mathcal{P}_{\mathcal R}(k)}\right) < \rho_{\rm PBH}^{\rm max}\,,
\ee 
where $a(M)$ corresponds to the horizon re-entry epoch, $k=a(M) H_M$, and the fraction $\beta$ of energy density in PBHs at $a(M)$ is given by Eq.~\eqref{betarad}.

\begin{figure}
\includegraphics[width=.45\textwidth]{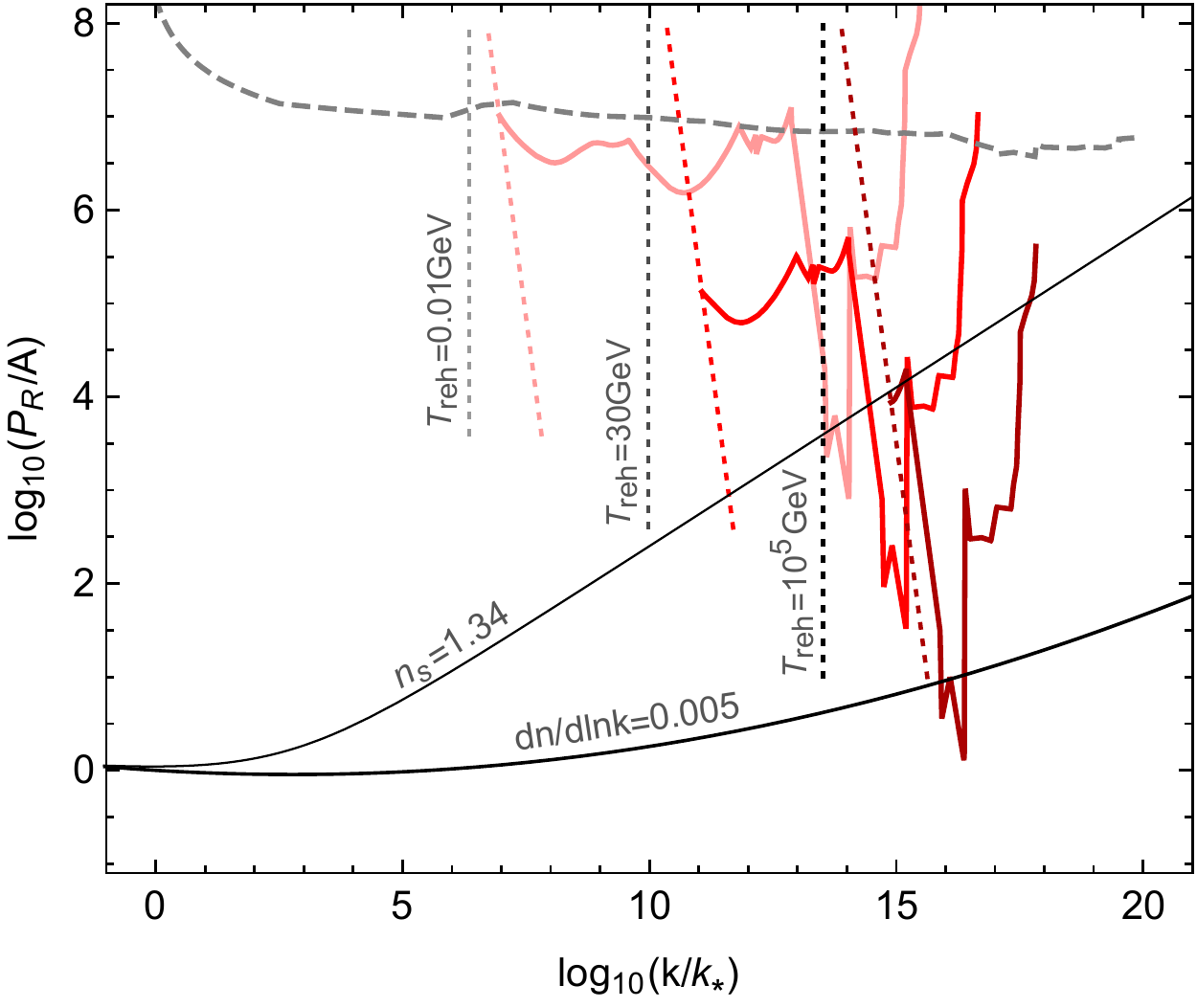}
\caption{Constraints on the amplitude of power spectrum. The gray dashed line corresponds to PBH production in a radiation-dominated era and the red lines in a matter-dominated era for reheating temperatures $10^5$ GeV, $30$ GeV and $0.01$ GeV from darkest to lightest. The red dashed lines show the smallest $k$ which become non-linear during the matter-dominated era, and the vertical black dashed lines the largest $k$ which cross the horizon after reheating. The black solid lines show the power spectrum with (thin) and without (thick) the spectator field. For the thin line the running of both the inflaton and spectator spectral indices are fixed to zero and the amplitude of the spectator field power spectrum to  $A_s=0.1A$. Here the constraints in the matter-dominated era are calculated without the factor $\sigma^{3/2}$. }
\label{PRk}
\end{figure}

The maximum $\mathcal{P}_{\mathcal R}(k)$ is shown in Fig.~\ref{PRk}. The gray dashed line shows the upper limit in the radiation-dominated case. The red lines show the limits in a matter-dominated case for different reheating temperatures, calculated assuming the matter dominance begins sufficiently early that the smallest constrained scales cross the horizon during matter dominance. These scales correspond to PBHs which evaporate during BBN. The left ends of the red lines correspond to the largest scales for which the perturbations become non-linear before the end of the matter-dominated era. The lowest peak in the red lines corresponds to the PBH constraint from CMB anisotropies associated with PBH evaporation~\cite{Carr:2009jm}\footnote{These constraints have been recently reanalysed in Ref.~\cite{Poulin:2016anj}. Although the data are significantly improved, the resulting constraints are very similar to the ones obtained in Ref.~\cite{Carr:2009jm} because the latter overestimated them. We use the constraints from Ref.~\cite{Carr:2009jm} because they are more easily represented.}.

We now use the constraints on the power spectrum for a monochromatic PBH mass function to constrain the running of the inflaton spectral index and the spectator scalar spectral index. The mass function is not monochromatic but the use of constraints for a monochromatic mass function is still justified in this case for the following reasons. First, the constraint for low reheating temperature, $T_{\rm reh}\lsim10^6\,{\rm GeV}$, arises from the CMB anisotropy and this is both dominant and very peaked. Thus the integral on the left-hand side of Eq.~\eqref{ineq} is dominated by the contribution at $M\simeq10^{-20}M_\odot$. Second, for reheating temperatures $T_{\rm reh}\gsim10^6\,{\rm GeV}$, the constraint arises from BBN. In that case, for the values of $A_s/A$ used, even a very small change in the running of the inflaton spectral index $\alpha$, or the spectator field spectral index $n_s$, implies a large change in the PBH abundance in the constrained mass region. Thus, assuming a monochromatic mass function causes only a very small error even for high reheating temperatures.

\subsection{Inflaton field with a running spectral index}

We first consider a scenario in which the inflaton perturbations determine the total power spectrum at all relevant scales. For example, the Universe may have been matter-dominated due to a prolonged period of reheating after inflation~\cite{Carr:1994ar,Allahverdi:2010xz}. In this case, we can study how PBH production constrains the running of the inflaton spectral index. The thick solid line in Fig.~\ref{PRk} illustrates the power spectrum for positive running, ${\rm d}n/{\rm d}\ln k=0.005$; this crosses the darkest red line and so $T_{\rm reh} = 10^5$\,GeV is excluded.

\begin{figure}
\includegraphics[width=.49\textwidth]{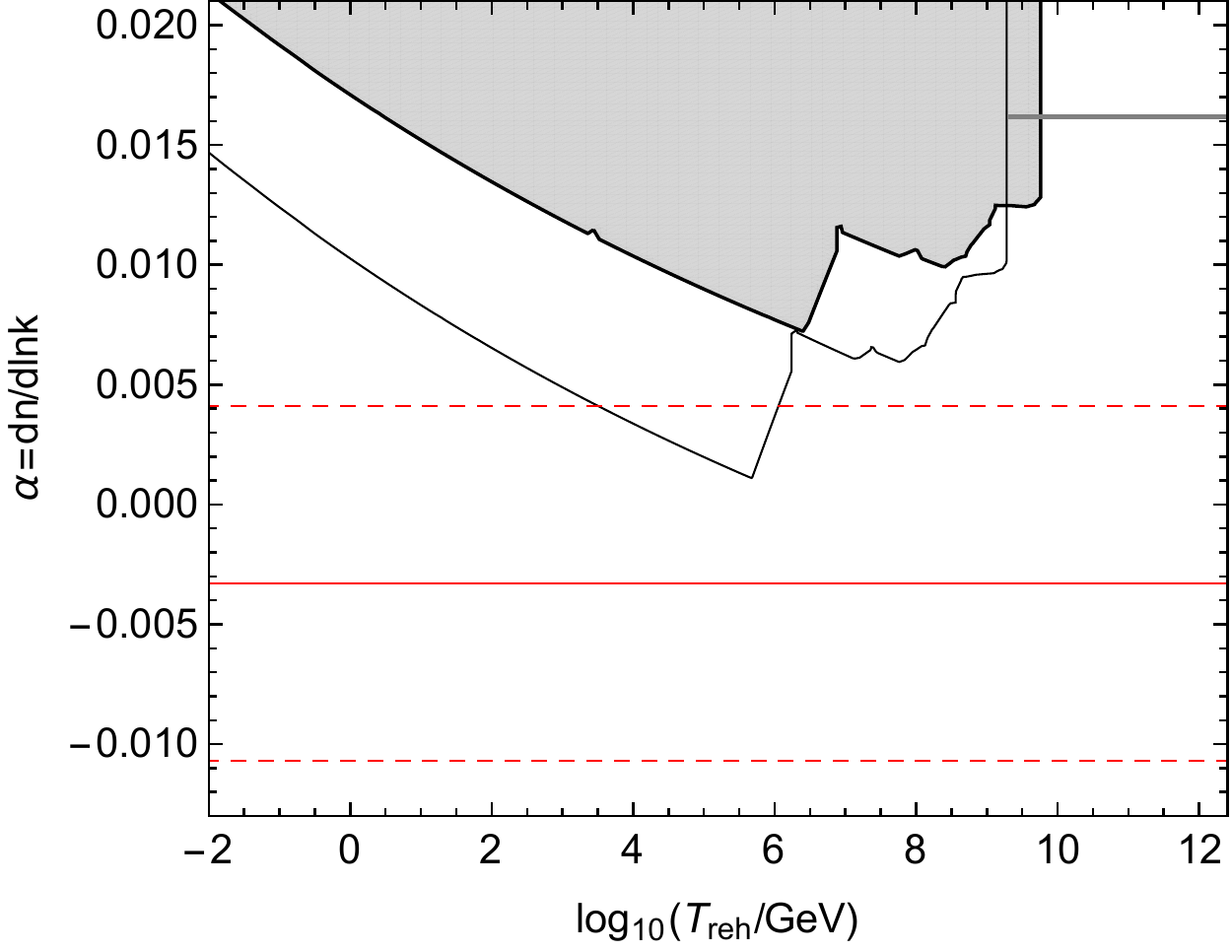} 
\caption{Constraints on the running of the inflaton spectral index. The thin black line shows the constraint as a function of the reheating temperature from PBH production during an early matter-dominated phase. The thick line shows the same constraint calculated with the extra factor $\sigma^{3/2}$. The horizontal gray line in the upper right corner shows the constraint from PBH production during radiation dominance. The red lines show the Planck result ${\rm d}n/{\rm d}\ln k = -0.0033\pm0.0074$.}
\label{running}
\end{figure}

In Fig.~\ref{running} the constraints on the running of the inflaton spectral index are shown. If there was an early matter-dominated era, the constraint on ${\rm d}n/{\rm d}\ln k$ depends on the reheat temperature. As in Fig.~\ref{PRk}, we have assumed that matter dominance begins before the smallest constrained scales cross the horizon. For $T_{\rm reh}\lsim 10^6$ GeV, the constraints arise from the CMB anisotropies and correspond to the lowest peaks of the red lines in Fig.~\ref{PRk}. For larger $T_{\rm reh}$, the constraints get weaker because the scale corresponding to the peak only crosses the horizon  after reheating and PBHs do not form at scales constrained by CMB anisotropies. The most stringent limit, ${\rm d}n/{\rm d}\ln k < 0.001$, which is comparable to the Planck limit \eqref{alpha}, is obtained for $T_{\rm reh}\sim 10^6$ GeV. The fact that even some negative values for the running of the inflaton spectral index are excluded demonstrates how efficient PBH formation is during a matter-dominated epoch.

As discussed in Sec.~\ref{formation}, we also show the constraints from PBH formation during an early matter-dominated era with the $\sigma^{3/2}$ factor in $\beta$. Obviously, the constraints become less stringent in this case because formation of PBHs becomes more suppressed. If the suppression factor $\sigma^{3/2}$ is taken into account, we find ${\rm d}n/{\rm d}\ln k < 0.007$ for $T_{\rm reh}\sim 10^6$ GeV.

For radiation dominance the upper limit on the running is ${\rm d}n/{\rm d}\ln k < 0.016$. This is also the constraint on the running for $T_{\rm reh} \gsim 10^{10}$ GeV, because in this case all PBHs produced during a matter-dominated era have evaporated before BBN. However, the running is still constrained by PBHs produced during the radiation-dominated era at $T<T_{\rm reh}$.

\begin{figure}
\includegraphics[width=.45\textwidth]{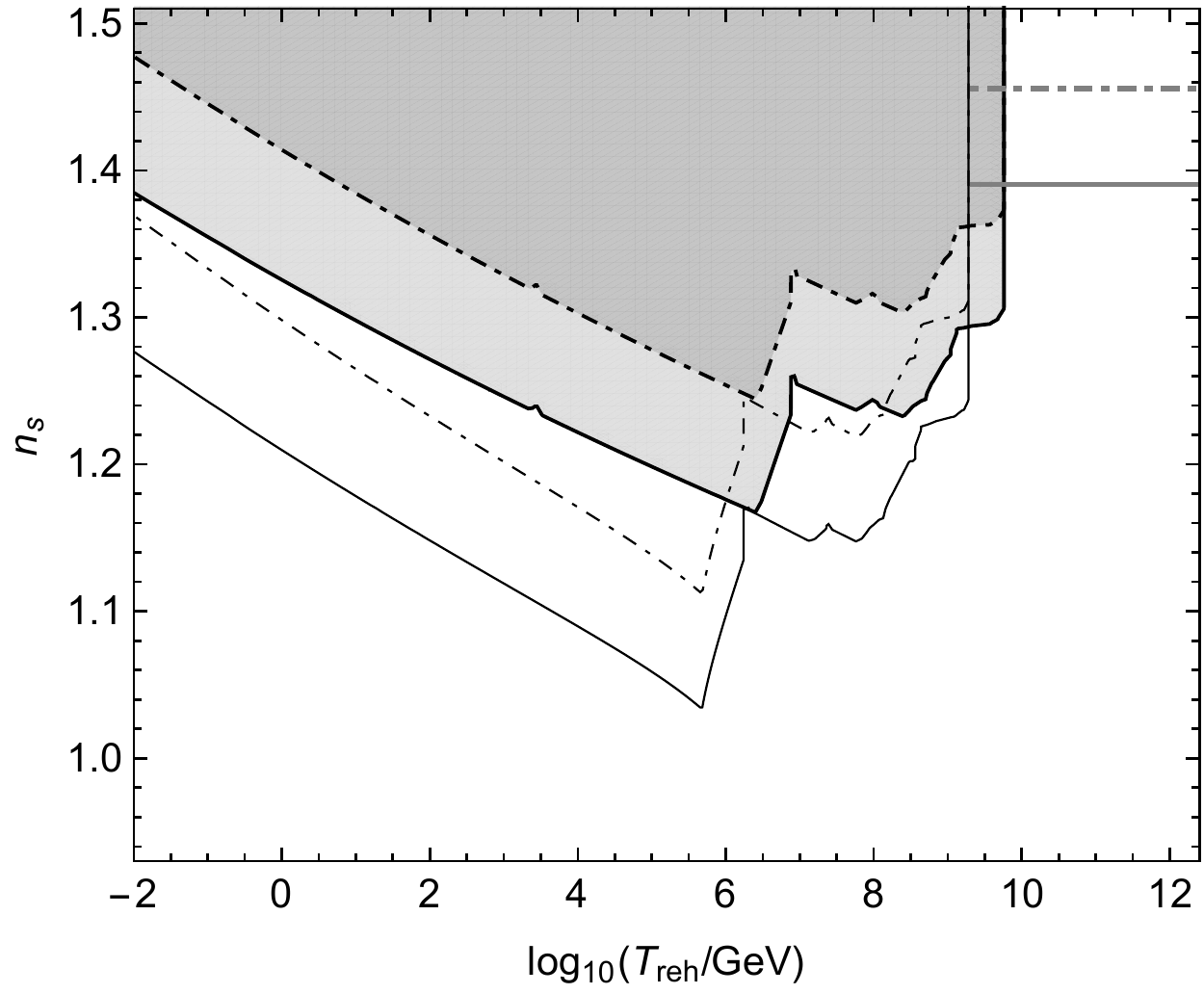} 
\caption{Constraints on the spectral index of a generic spectator field. The thin black lines show the constraint as a function of the reheating temperature arising from PBH production during an early matter-dominated phase for $A_s=0.1A$ (solid line) and $A_s=0.005A$ (dashed line). The thick lines show the same constraints calculated with the factor $\sigma^{3/2}$. The horizontal gray lines in the upper right corner show the constraint arising from PBH production during radiation dominance.}
\label{spectator}
\end{figure}

\subsection{Generic spectator field}

Finally, we consider the scenario introduced in Sec.~\ref{spectrum}, where the total power spectrum is dominated by spectator field perturbations at large $k$. This is illustrated by the thin solid line in Fig.~\ref{PRk}. The constraints on the spectator field spectral index $n_s$ for different amplitudes $A_s$ are shown in Fig.~\ref{spectator}. For simplicity we have neglected the running of both the inflaton and spectator field spectral indices. The constraints can be understood as in the previous case. In the absence of a matter-dominated era, the constraint on $n_s$ from PBH formation during a radiation-dominated era is
\be
 n_s < 1.34 - 0.02\log_{10}(A_s/A) \, .
\ee

\section{Conclusions}
\label{conclusions}
We have studied the production of PBHs during an early matter-dominated phase. We have considered two possible sources of perturbations: a spectator field with a blue spectrum, which allows significant PBH production at small scales, and an inflaton field with a running spectral index. If the running of spectral indices is small, the resulting PBH mass function has a power law form characterised by cut-offs at $M_{\rm min}$ and $M_{\rm max}$, related to the times when the matter-dominated period starts and ends.

We have identified the region of parameter space where a significant fraction of the DM can be produced, taking into account all current PBH constraints. Whether one can obtain all the DM with a nearly flat mass function depends on the validity of dynamical constraints. We have also presented constraints on the amplitude and spectral index of the spectator field as a function of the reheating temperature and on the running of the inflaton spectral index, ${\rm d}n/{\rm d}{\rm ln}k \lesssim -0.002$. The latter is comparable to that from the Planck satellite for a scenario where the spectator field is absent.

\vspace{4mm}
\noindent
\textbf{Note added in proof:} After the first version of this paper appeared, Ref.~\cite{Cole:2017gle} pointed out that our expression for $\sigma(k)$ was missing a factor of 1/3. This has now been included.

\section*{Acknowledgements}
We thank C. Byrnes, S. Camera, P. Cole, S. Galli, H. Kurki-Suonio, A. Kusenko, K. Malik, D. Mulryne, A. Polnarev, M. Raidal, J. V\"aliviita,  H. Veerm\"ae and an anonymous referee for correspondence and discussions. T.T. is supported by the U.K. Science and Technology Facilities Council grant ST/J001546/1 and V.V. by the Estonian Research Council grant IUT23-6 and ERDF Centre of Excellence project No TK133.

\bibliography{mdom.bib}

\end{document}